\documentclass[longauth]{aa}

\usepackage{graphicx}
\usepackage{txfonts}

\usepackage{amssymb}
\usepackage{latexsym}
\usepackage{amstext}
\usepackage{natbib}
\bibpunct{(}{)}{;}{a}{}{,}

\newcommand{\vsini}{$\mathrm{v}\sin{i}$}
\newcommand{\Brg}{$\mathrm{Br}\gamma$}

\newcommand{\degrees}{\ensuremath{^{\circ}}}

\newcommand{\luan}{4}
\newcommand{\laog}{2}
\newcommand{\oca}{3}
\newcommand{\mpifr}{5}
\newcommand{\oaa}{1}

\begin{document}
\title{Constraining the wind launching region in Herbig Ae stars: 
AMBER/VLTI spectroscopy of  HD104237}
%
%
%
\author{E.~Tatulli\inst{\oaa}                                 
   \and I.~Isella\inst{\oaa,12}
  \and A.~Natta\inst{\oaa}
  \and L.~Testi\inst{\oaa}   
  \and A.~Marconi\inst{\oaa}                           
  \and F.~Malbet\inst{\laog}
  \and P.~Stee\inst{\oca}                                       
  \and    R.G.~Petrov\inst{\luan}                  
  \and F.~Millour\inst{\luan,\laog}             
  \and A.~Chelli\inst{\laog}                  
  \and G.~Duvert\inst{\laog}                    
\and P.~Antonelli\inst{\oca}                  
  \and U.~Beckmann\inst{\mpifr}                 
  \and Y.~Bresson\inst{\oca}                    
  \and M.~Dugu\'e\inst{\oca}                  
  \and L.~Gl\"uck\inst{\laog}                   
  \and P.~Kern\inst{\laog}                      
  \and S.~Lagarde\inst{\oca}                    
  \and E.~Le Coarer\inst{\laog}                 
  \and F.~Lisi\inst{\oaa}                       
  \and K.~Perraut\inst{\laog}                   
  \and P.~Puget\inst{\laog}                     
  \and S.~Robbe-Dubois\inst{\luan}              
  \and A.~Roussel\inst{\oca}
  \and G.~Weigelt\inst{\mpifr}                    
  \and G.~Zins\inst{\laog}                        
  \and M.~Accardo\inst{\oaa}                                    
  \and B.~Acke\inst{\laog,13}                                   
  \and K.~Agabi\inst{\luan}                                   	  
  \and B.~Arezki\inst{\laog}                                            
  \and E.~Aristidi\inst{\luan}                                  
  \and E.~Altariba\inst{\luan}                         
  \and C.~Baffa\inst{\oaa}                                        
  \and J.~Behrend\inst{\mpifr}                                          
  \and T.~Bl\"ocker\inst{\mpifr}                                  
  \and S.~Bonhomme\inst{\oca}                                           
  \and S.~Busoni\inst{\oaa}                                       
  \and F.~Cassaing\inst{6}                                              
  \and J.-M.~Clausse\inst{\oca}                        	          
  \and J.~Colin\inst{\oca}		          
  \and C.~Connot\inst{\mpifr}                          
  \and A.~Delboulb\'e\inst{\laog}                      
  \and A.~Domicinano de Souza\inst{\luan}              
  \and T.~Driebe\inst{\mpifr}                          	    
  \and P.~Feautrier\inst{\laog}                        		  
  \and D.~Ferruzzi\inst{\oaa}                          	  
  \and T.~Forveille\inst{\laog}                        		  
  \and E.~Fossat\inst{\luan}                           	          
  \and R.~Foy\inst{7}                                  		  
  \and D.~Fraix-Burnet\inst{\laog} 
  \and A.~Gallardo\inst{\laog}                         	  
  \and S.~Gennari\inst{\oaa}                           	  
  \and A.~Glentzlin\inst{\oca}                         		  
  \and E.~Giani\inst{\oaa}                             		  
  \and C.~Gil\inst{\laog,14}                           		  
  \and M.~Heiden\inst{\mpifr}                          	          
  \and M.~Heininger\inst{\mpifr}                       		  
  \and O.~Hernandez Utrera\inst{\laog}                   
  \and K.-H.~Hofmann\inst{\mpifr}                        
  \and D.~Kamm\inst{\oca}                                
  \and S.~Kraus\inst{\mpifr}                         
  \and D.~Le Contel\inst{\oca}                         
  \and J.-M.~Le Contel\inst{\oca}                      
  \and B.~Lopez\inst{\oca}                               
  \and Y.~Magnard\inst{\laog}                            
  \and G.~Mars\inst{\oca}                              
  \and G.~Martinot-Lagarde\inst{8,15}                    
  \and P.~Mathias\inst{\oca}                           
  \and J.-L.~Monin\inst{\laog}                           
  \and D.~Mouillet\inst{\laog,16}                      
  \and D.~Mourard\inst{\oca}                             
  \and P.~M\`ege\inst{\laog}                           
  \and E.~Nussbaum\inst{\mpifr}                          
  \and K.~Ohnaka\inst{\mpifr}                        
  \and J.~Pacheco\inst{\oca}		          
  \and C.~Perrier\inst{\laog}                        	            
  \and Y.~Rabbia\inst{\oca}                                          
  \and S.~Rebattu\inst{\oca}                         	          
  \and F.~Reynaud\inst{9}                            	            
  \and A.~Richichi\inst{10}                          	            
  \and M.~Sacchettini\inst{\laog}                           	  
  \and P.~Salinari\inst{\oaa}                                           
  \and D.~Schertl\inst{\mpifr}                                          
  \and W.~Solscheid\inst{\mpifr}                                
  \and P.~Stefanini\inst{\oaa}                                          
  \and M.~Tallon\inst{7}                             	    
  \and I.~Tallon-Bosc\inst{7}                               	    
  \and D.~Tasso\inst{\oca}                                                      \and J.-C.~Valtier\inst{\oca}                                         
  \and M.~Vannier\inst{\luan,11}                            	  
  \and N.~Ventura\inst{\laog}                        		
  \and M.~Kiekebusch\inst{11}                                     
  \and M.~Sch\"oller\inst{11}
}             

\institute{
  INAF-Osservatorio Astrofisico di Arcetri, Istituto Nazionale di
  Astrofisica, Largo E. Fermi 5, I-50125 Firenze, Italy
  \and Laboratoire d'Astrophysique de Grenoble, UMR 5571 Universit\'e Joseph
  Fourier/CNRS, BP 53, F-38041 Grenoble Cedex 9, France
  \and Laboratoire Gemini, UMR 6203 Observatoire de la C\^ote
  d'Azur/CNRS, BP 4229, F-06304 Nice Cedex 4, France
  \and Laboratoire Universitaire d'Astrophysique de Nice, UMR 6525
  Universit\'e de Nice/CNRS, Parc Valrose, F-06108 Nice cedex 2,
  France
  \and Max-Planck-Institut f\"ur Radioastronomie, Auf dem H\"ugel 69,
  D-53121 Bonn, Germany
  \and ONERA/DOTA, 29 av de la Division Leclerc, BP 72, F-92322
  Chatillon Cedex, France 
  \and Centre de Recherche Astronomique de Lyon, UMR 5574 Universit\'e
  Claude Bernard/CNRS, 9 avenue Charles Andr\'e, F-69561 Saint Genis
  Laval cedex, France
  \and Division Technique INSU/CNRS UPS 855, 1 place Aristide
  Briand, F-92195 Meudon cedex, France
  \and IRCOM, UMR 6615 Universit\'e de Limoges/CNRS, 123 avenue Albert
  Thomas, F-87060 Limoges cedex, France
  \and European Southern Observatory, Karl Schwarzschild Strasse 2,
  D-85748 Garching, Germany
  \and European Southern Observatory, Casilla 19001, Santiago 19,
  Chile
  \and Dipartimento di Fisica, Universita di Milano, via Celoria 16, 20133 Milano, Italy
  \and Instituut voor Sterrenkunde, KULeuven, Celestijnenlaan 200B,
  B-3001 Leuven, Belgium 
  \and Centro  de  Astrof\'{\i}sica  da  Universidade  do  Porto, Rua
  das Estrelas - 4150-762 Porto, Portugal 
  \and \emph{Present affiliation:} Observatoire de la Côte d'Azur -
  Calern, 2130 Route de l'Observatoire , F-06460 Caussols, France
  \and \emph{Present affiliation:} Laboratoire Astrophysique de
  Toulouse, UMR 5572 Universit\'e Paul Sabatier/CNRS, BP 826, F-65008
  Tarbes cedex, France 
}

\offprints{E.~Tatulli\\ email: \texttt{<etatulli@arcetri.astro.it>}}

\date{Received date; accepted date}
\abstract{}{We investigate the origin of the $\mathrm{Br}\gamma$ emission 
  of the Herbig Ae star HD104237 on Astronomical Unit (AU) scales.} 
  {Using AMBER/VLTI  at a spectral resolution $\mathcal{R}=1500$  we 
  spatially resolve the emission in both the $\mathrm{Br}\gamma$ line 
  and the adjacent continuum.} 
  {The visibility does not vary between the continuum and the \Brg\ line, 
  even though the line is strongly detected in the spectrum, with a peak 
  intensity 35\% above the continuum. This demonstrates that the line and
  continuum emission have similar size scales. We assume that the  K-band
  continuum excess originates in a ``puffed-up'' inner rim of the 
  circumstellar disk, and discuss the likely origin of \Brg.}
  {We conclude that this emission most likely arises from a compact  
  disk wind, launched from a region 0.2-0.5 AU from the star, with a 
  spatial extent similar to that of the near infrared continuum  
  emission region, i.e, very close to the inner rim location.}  

\maketitle
%
\section{Introduction}

The spectra of pre-main sequence stars of all masses show prominent strong 
and broad emission lines, of both hydrogen and metals. These lines trace the
complex circumstellar environment which characterizes this evolutionary
phase, and are very likely powered by the associated accretion disks. 
The emission lines are used to infer the physical properties of the
gas, and to constrain its geometry and dynamics. Their exact origin,
however, is not known. The hydrogen lines, in particular, may
originate either in the  gas which accretes onto the star from the disk,
as in magnetospheric accretion models \citep{hartmann_1}, or in  winds 
and jets, driven by the interaction of the accreting disk with a stellar 
\citep{shu_1} or disk \citep{casse_1} magnetic field.  For Herbig Ae stars, 
it is additionally possible that they form in the inner disk \citep{tb}. 

\begin{figure*}
\begin{center}
\includegraphics[height=5cm, angle=0]{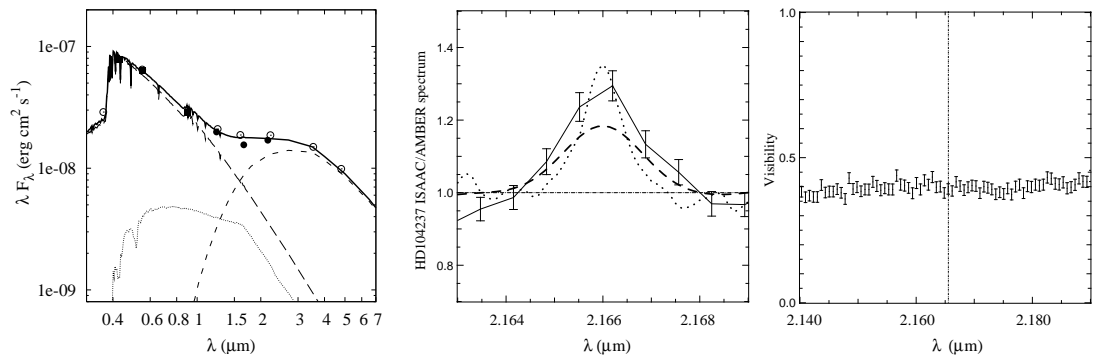}
\caption{\label{fig_visHD104237}Left:  SED predicted by the ``puffed-up'' 
rim model used to normalise the K-band continuum visibility. The
contributions of the A star (long-dashed line), the K3
star (dotted line) and the rim (short dashed line) are shown  (see the
text for the stellar parameters). The observations are from 2MASS
(filled circles), Hipparcos (filled squares), and \citet{malfait_1} (empty
circles).  An extinction of $\mathrm{A_v} = 0.31$ \citep{ancker_1} has been used to correct the data for interstellar absorption. Center: Comparison of 
$\mathrm{Br}\gamma$  observed with AMBER in the photometric channels (solid 
line) and ISAAC  (dotted line); the dashed line shows the ISAAC spectrum
smoothed to the spectral resolution of AMBER.  Right: Visibility of HD104237 
as a function of wavelength. The continuum has been normalised using the 
star~+~rim model, as described in the text. The vertical line shows the 
$\mathrm{Br}\gamma$ wavelength.
}
\end{center}
\end{figure*}

For all models, emission in the hydrogen lines is predicted to occur 
over very small spatial scales, a few AUs at most. To understand the 
physical processes that occur at these scales, one needs to combine 
very high spatial resolution with enough spectral resolution to resolve
the line profile. AMBER, the three-beam near-IR recombiner of the VLTI 
\citep{petrov_1}, simultaneously offers high spatial and high spectral 
resolution, with the sensitivity required to observe pre-main sequence 
stars. These capabilities were recently demonstrated by \citet{malbet_1}, 
who succesfully resolved the luminous Herbig Be system MWC297 in both the 
continuum and the $\mathrm{Br}\gamma$ line. They showed that the line
emission originates from an extended wind, while the continuum infrared 
excess traces a dusty accretion disk. 

In this Letter we concentrate on a lower mass, less active system, the
Herbig Ae system HD104237. The central emission line star, of
spectral type between A4V and A8,  is
surrounded by a circumstellar disk, which causes the infrared excess
emission \citep{meeus_1} and drives a jet seen in Ly-$\alpha$ images
\citep{grady_1}. The optical spectrum shows a rather narrow H$\alpha$ 
emission with a P-Cygni profile \citep{feigelson_1}. The disk is seen
almost pole-on ($i=18^{\degrees +14}_{~-11}$; Grady et
al.~\citeyear{grady_1}), consistent with the low value of \vsini\ (12
km s$^{-1}$; Donati et al.~\citeyear{donati_1}). \citet{donati_1} 
detected a stellar magnetic field of 50 G. \citet{bohm_1} have
revealed the presence of a very close companion of spectral type K3,  orbiting with a period of
$\sim 20$ days, and which bolometric luminosity is $10$ times fainter than the one of the central star. In the near infrared
domain, spatially unresolved ISAAC observations (Garcia Lopez et al.,
2006) show a strong $\mathrm{Br}\gamma$ emission line, with a peak flux 
35\% above the continuum.

\section{Observations and data reduction}

AMBER observed HD104237 on 26 February 2004 on the UT2-UT3 baseline of 
the VLTI, which corresponds to a projected length of $B = 35\mathrm{m}$. 
The instrument was set up to cover the $[2.121, 2,211]\mu\mathrm{m}$ 
spectral range with a spectral resolution of $\mathcal{R} = 1500$, 
which resolved the profile of the $\mathrm{Br}\gamma$ emission line at 
$2.165\mu\mathrm{m}$. The data consist of $10$ exposures of $500$ frames, 
with an integration time of $100\mathrm{ms}$ for each frame. The integration 
time is a trade-off between gathering enough flux in the photometric channel 
and preventing excessive contrast loss due to fringe motion during the
integration. At the time of these early observations the star magnitude of 
K=$4.6$ was close to the sensitivity limit of the medium spectral resolution 
mode, and as a result fringes are only visible in $\approx 25\%$ of the 
frames. In the photometric channels the $\mathrm{Br}\gamma$ line
is under the noise level of the indivual frames, but it is well detected on 
their average and contributes $35\%$ of the total flux 
(Fig.~\ref{fig_visHD104237}, central panel). For comparison we overlay
a higher spectral resolution spectrum ($\mathcal{R} = 8900$, taken with
ISAAC approximately one year before the VLTI observations), as well as 
its smoothing to the AMBER spectral resolution. The two spectra are 
consistent with each other, within the combined typical line variability 
of Herbig Ae stars and calibration uncertainties.

Data reduction followed standard AMBER procedures \citep{tatulli_1}. 
To optimize the signal to noise ratio (SNR) of the visibility derived
from these relatively poor-quality data, we pre-selected frames with 
individual SNR greater than $1.5$, ensuring that fringes are present
in all short exposures that enter the visibility.    

For these observations the visibility amplitude could not be calibrated 
to an absolute scale using an astronomical calibrator (unresolved star),
due to non-stationary vibrations in the optical train of the UT telescopes. 
With a 100ms integration time these vibrations randomly degrade the 
fringe contrast of the individual frames by a large factor, and the statistics
of that attenuation depends on uncontrolled factors (exact telescope 
and delay line pointing, recent environmental history, etc) that do not 
similarly affect the target star and its calibrator. Fortunately the
contrast loss from vibrations is achromatic across our small relative
bandpass, and as a consequence the differential visibility, that is, the relative visibility between the
spectral channels, is unaffected. Our dataset therefore allows to 
investigate the visibility across the $\mathrm{Br}\gamma$ emission 
line compared to that in the adjacent continuum. The  differential
visibilities are accurate to approximately 5\%. 

The excess near infrared continuum emission in Herbig Ae stars most
likely originates in the innermost regions of their dusty circumstellar 
disks, at the dust sublimation radius \citep{eisner_2,isella_2}. To 
approximately normalise the visibilities, we thus scaled the
observed continuum value to the predictions of appropriate theoretical
models of the disk inner rim emission \citep{isella_1}. We computed the 
structure of the ``puffed-up'' inner rim as described in 
\citet{isella_2}. We used for the two stars the following parameters:
$T_{eff}=8000K$, $L_{\star}=30L_{\sun}$ for the primary and
$T_{eff}=4700K$, $L_{\star}=3L_{\sun}$ for the secondary star, respectively. We adopted a
distance of $D=115pc$, and a disk mass surface density of
$\Sigma(r)=2\cdot 10^3 \cdot r^{-3/2}$g/cm$^2$ (with $r$ in AU).

Fig. \ref{fig_visHD104237} (left) shows that the model using micron size   
astronomical silicates produces a very good fit of the SED. The star 
fluxes contribute approximately $30\%$ (respectively $20\%$ and
$10\%$ for the primary and secondary star, which gives a binary flux ratio of $\sim0.5$ in the K band) of the total $2.15\mu\mathrm{m}$ 
flux, and the inner rim appears as a bright ring with radius
$R_{rim}=0.45\mathrm{AU}$ ($3.8\mathrm{mas}$ at $D=115pc$). 
The resulting model visibility on the 35m baseline is $V=0.38$ 
(Fig~\ref{fig_visHD104237}, right). Note that the orbital period
of the spectroscopic binary is $~20$ days. This leads to an average
separation of $\delta{s} = 0.15\mathrm{AU}$ which corresponds to an
angular separation of $1.2\mathrm{mas}$. The central system (A star + K3 companion)  is therefore  completely unresolved on the current baseline, and both stars are located inside the dust evaporation radius.

\section{Results and discussions}

\begin{figure}[t]
\begin{center}
  \includegraphics[width=12cm,angle=270]{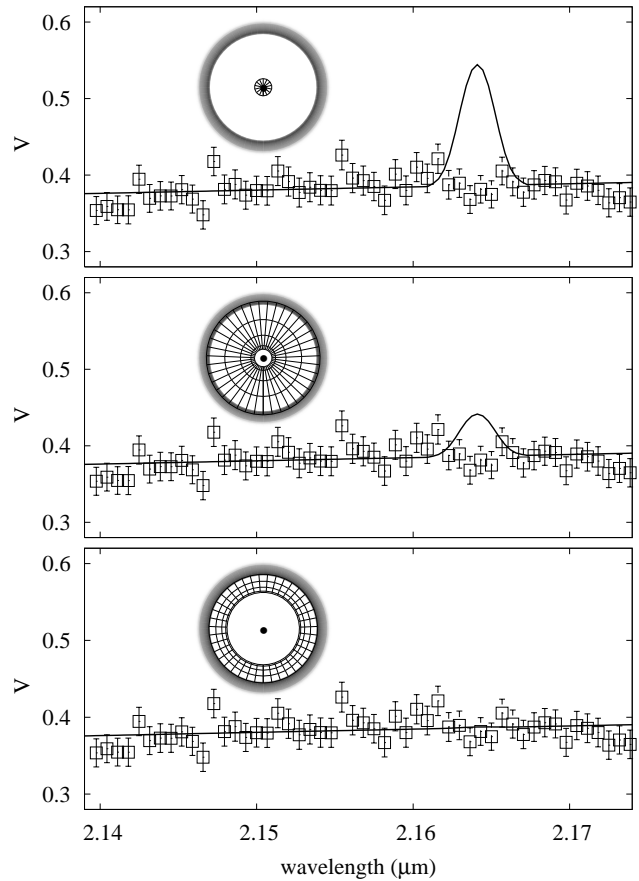} 
\caption{\label{fig_models} Comparison between the observed visibilities
  (empty square with error bars) and the predictions
  (solid curves) of the simple geometrical models for the
  $\mathrm{Br}\gamma$ emission (sketched in the same panels). 
  The observed visibilities are scaled to match the continuum value 
  predicted by the ``puffed-up'' inner rim model (Isella \& Natta, 2005) 
  as described in Sec.2. The continuum emission arises both from the
  stellar photosphere (${\approx}20\%$) and from the dusty disk inner
  rim, located at the dust evaporation distance $R_{rim}=0.45$AU and
  which appears as the  bright gray scale ring. 
  The $\mathrm{Br}\gamma$ emission regions are shown as 
  grid surfaces. The three panels illustrate the different models 
  discussed in Sec.3: the upper panel represents the {\it magnetospheric
  accretion} model in which the $\mathrm{Br}\gamma$ emission
  originates very close to the star, inside the corotational
  radius $R_{corot}=$0.07AU; in the middle panel the $\mathrm{Br}\gamma$ 
  emission originates between $R_{corot}$  and the rim radius
  $R_{rim}=$0.45AU, representing the {\it gas within the disk}
  model; the bottom panel shows the {\it outflowing wind} model,
  in which the emission is confined close to the inner
  rim, between $\sim0.2$AU and $\sim0.5$AU.}   
\end{center}
\end{figure}

  Within the fairly small error bars, the visibility does not 
  change across the  $\mathrm{Br}\gamma$ emission line. This result
  is robust and puts strong constraints on the relative spatial 
  extent of the line and continuum emission regions, demonstrating 
  that they have very similar apparent sizes. We use this constraint 
  to probe the processes responsible for the the $\mathrm{Br}\gamma$
  emission in this star, and consider in turn the three main
  mechanisms usually invoked to interpret the hydrogen line emission
  in pre-main-sequence stars. We translate each mechanism to a simple 
  geometrical models of specific spatial extension, with the line 
  strength fixed at the observational value, and evaluate the resulting
  visibility across the line\footnote{Note that this analysis is valid as long as the continuum emission is {\it resolved}. In our case the continuum is calibrated by a model and not by an unresolved star, and therefore there might be a chance that the visibility of the continuum is close to $1$. However, this peculiar scenario appears very unlikey since the puffed-up rim model has been shown to be well representative of the very close environment of Herbig Ae stars over a large luminosity range \citep{monnier_1}. Furthermore, the hypothesis of a resolved continuum emission for HD104237 is  strongly supported  by its measured accretion rate of $10^{-8}M_{\sun}$ \citep{grady_1}. Indeed, \citet{muzerolle_1} showed that for weak accretion rates ($\dot{M} < 10^{-7}M_{\sun}/\mathrm{yr}$), the gaz accreting onto the star is optically thin. It therefore does not shield the dust from the stellar radiation and consequently, the star must exhibit an inner region free of dust, large enough to be resolved by the interferometer at the $35\mathrm{m}$ baseline.}.

{\it Magnetospheric accretion:} in such a model matter infalls
on the star along magnetic field lines, and the base of this infalling
flow is (approximately) inside the corotation radius \citep{muzerolle_1}. 
For HD104237 we find  $R_{corot} =  0.07\mathrm{AU}$, using 
${v}\sin{i}=12\mathrm{km}/\mathrm{s}$ \citet{donati_1} and an inclination of $i=18^{\degrees}$ \citep{grady_1}. Note that this value is, at our spatial resolution level,  similar to the separation of the binary. Adopting $R_{star}$ and $R_{corot}$  as the inner and outer limits of the 
magnetospheric accreting region, Fig. \ref{fig_models} (upper panel) 
demonstrates that $\mathrm{Br}\gamma$ emission is then confined much 
closer to the star than the dusty rim. The predicted visibility 
therefore increases significantly in the line, contrary to the observational
result. Explaining the observed visibility with magnetospheric accretion
requires the corotation radius to approach the inner rim, which would
need an unrealistically lower  stellar rotational velocity ($\mathrm{v}<
2\mathrm{km}/\mathrm{s}$). The magnetic field of HD104237 is
weak, $B=50G$ \citep{donati_1}, and the corresponding magnetospheric 
truncation radius ($R_{mag} = 0.018\mathrm{AU}$, \citet{shu_1}) is 
smaller than the corotation radius. Replacing $R_{corot}$ by $R_{mag}$ 
as the outer radius of the magnetospheric accreting region would thus 
only reinforce our conclusion. Magnetospheric accretion therefore cannot 
be responsible for most of the $\mathrm{Br}\gamma$ emission. 

{\it Gas in the disk:} since HD104237 accretes matter (a few 
$10^{-8}M_{\sun}/\mathrm{yr}$, Grady et al.~\citeyear{grady_1}, 
Garcia Lopez et al.~\citeyear{garcia_1}), an optically thin gaseous 
disk (Muzerolle et al. 2004) must extend inward from the dust evaporation
rim to either the magnetospheric truncation radius or the corotation 
radius; this disk region is ionised and may, in principle, emit the 
observed \Brg\ line. To model this scenario, we assume a constant 
$\mathrm{Br}\gamma$ surface brightness between $R_{corot}$ (or $R_{mag}$) 
and $R_{rim}$. The resulting visibilities (Fig. \ref{fig_models}, middle 
panel) are also inconsistent with the observations, though not quite
as severely. 

{\it Outflowing wind:} the remaining possibility is that 
$\mathrm{Br}\gamma$ is emitted in  a wind or jet. The bulk of 
the emission is then confined to the regions of highest gas density, 
i.e., to the base of  the wind/jet. Since the wind/jet must be seen 
almost pole-on, we assume in our toy model that emission is confined 
to a ring of width $\Delta R$, with  $\Delta R/R_i \sim  0.5$, where 
$R_i$ is the ring inner radius, that is the wind launching point. This 
assumption is guided by the wind models  of  \citet{natta_2}, and by
the more recent  simulations  of the $\mathrm{Pa}\beta$ emission
in jets of T Tauri stars \citep{thiebaut_1}. For pole-on outflows
most of the intensity originates in a ring with these approximate
properties. Fixing  $\Delta R/R_i = 0.5$ and adjusting $R_i$, this 
toy model reproduces the interferometric data for $R_i$ between
$\sim 0.25~\mathrm{AU}$ and $0.35~\mathrm{AU}$ (Fig. \ref{fig_models}, 
bottom panel). Allowing different relative widths ($\Delta R$ from $10\%$ 
to $100\%$ of $R_i$), we find  that the \Brg\ emission has to originate
between $\sim 0.2$ and $0.5$ AU. Our interferometric measurements are 
therefore consistent with the $\mathrm{Br}\gamma$ emission coming from 
the base of a  wind, originating in a disk region close to the location 
of the dusty rim. The presence of a wind in HD104237 is well established
on (much) larger  scales, by the  $\mathrm{Ly}\alpha$ bipolar microjet 
\citep{grady_1} and by the P-Cygni profile of the $\mathrm{H}\alpha$  
line \citep{feigelson_1}. Given our oversimplified geometrical models, it 
would be inappropriate to speculate too much on the detailed nature of the
wind. We note however that the launching region of an X-wind, driven
by the stellar magnetic field (Shu et al. 1994), is  close to the
corotation radius and too small to be consistent with the present 
data. Making the X-wind acceptable needs relaxing the assumption that 
most of the line emission originates near the launching point, and
instead having the brightest  \Brg\ region a factor $5-8$ further away 
from the star. The disk-wind scenario by contrast has its launching point a
few tenths of AU from the star and needs no tweaking. This preference
for a disk-wind	 is consistent with the size of jet launching regions 
inferred from the rotation of TTauri jets \citep{coffey_1}. 

It is important to recall here that our analysis assumes that the 
continuum originates in a dusty ``puffed-up'' ring. That model has 
been shown to match Herbig Ae and T Tauri stars over a large luminosity
range \citep{monnier_1}, but it obviously needs  to be verified for the 
specific case of HD104237, by obtaining calibrated interferometric 
observations over a few baselines.


\section{Conclusions}\label{sec_conc}

We have presented interferometric observations of the
Herbig Ae star HD104237, obtained with the AMBER/VLTI
instrument with $\mathcal{R}=1500$ high spectral resolution.
The visibility is identical in the $\mathrm{Br}\gamma$
line and in the continuum, even though the line represents $35\%$ of 
the total $2.165\mu\mathrm{m}$ flux. This immediately implies that 
the line and continuum emission regions have the same apparent size.

Scaling the continuum visibility with a ``puffed-up'' inner rim  model, 
and using simple toy models to describe  the  $\mathrm{Br}\gamma$
emission, we have shown that the line emission is unlikely to originate
in either magnetospheric accreting columns of gas or in the gaseous disk. 
It is much more likely to come from a compact outflowing disk wind  
launched in the vicinity of the rim, about 0.5 AU from the star. 
This does not preclude accretion from nonetheless occuring along the
stellar magnetic field detected by \citet{donati_1}, and accreting
matter might even dominate the optical hydrogen line emission, but
our observations show that the bulk of the \Brg\ emission in HD104237 
is unlikely to originate in magnetospheric accreting matter.

Our results show that AMBER/VLTI is a powerful diagnostic of
the origin of the line emission in young stellar objects. Observations 
of a consistent sample of objects will strongly constrain the wind 
launching mechanism.

\begin{acknowledgements}
  This work has been partly supported by the MIUR COFIN grant 2003/027003-001 and 025227/2004 to the INAF-Osservatorio Astrofisico di Arcetri. This project has benefited from funding from the French Centre
  National de la Recherche Scientifique (CNRS) through the Institut
  National des Sciences de l'Univers (INSU) and its Programmes
  Nationaux (ASHRA, PNPS). The authors from the French laboratories
  would like to thank the successive directors of the INSU/CNRS
  directors. C. Gil  work was supported in part by the Funda\c{c}\~ao para a Ci\^encia e a Tecnologia through project POCTI/CTE-AST/55691/2004 from POCTI,with funds from the European program FEDER. 

  This work is based on observations made with the European Southern
  Observatory telescopes. This research has also made use of
  the ASPRO observation preparation tool from the \emph{Jean-Marie
    Mariotti Center} in France, the SIMBAD database at CDS, Strasbourg
  (France) and the Smithsonian/NASA Astrophysics Data System (ADS).
  The data reduction software \texttt{amdlib} is freely available on
  the AMBER site \texttt{http://amber.obs.ujf-grenoble.fr}. It has
  been linked with the free software
  Yorick\footnote{\texttt{ftp://ftp-icf.llnl.gov/pub/Yorick}} to
  provide the user friendly interface \texttt{ammyorick}.
\end{acknowledgements}


\bibliographystyle{aa}

\end{document}